\documentclass[aps,prl,twocolumn,showpacs,superscriptaddress,groupedaddress]{revtex4}  
\usepackage{graphicx}  
\usepackage{dcolumn}   
\usepackage{bm}        
\usepackage{amssymb}   

\begin{document}

{\bf Reply to the Comment on \textquotedblleft Unified Formalism of Andreev Reflection at a Ferromagnet/Superconductor Interface" by Eschrig et al.}

	After the publication of the paper by Chen, Tesanovic, and Chien \cite {chen1} (hereafter noted as the CTC model) on the unified formalism of Andreev reflection, Eschrig et al.\cite {schrig2} (hereafter noted as the Comment), assert that a number of works [2-7 in Ref.2] have already solved the problem and that the CTC model violates basic physics.  Both assertions are false.

	In metals, the spin polarization $P$, where $0\leq P\leq 1$, is a quantity intrinsic to the material and uniquely defined by its band structure and carrier density. The Andreev reflection spectroscopy (ARS) at the interface between a normal metal (N) and a superconductor (S), can measure $P$ of the normal metal and the gap of the superconductor through proper analyses, therefore is of great importance. Over the last three decades, there have been many theoretical models pertaining to ARS, but only some are relevant to, and confirmed by, the ARS experiments. Numerous experimental ARS measurements using point contact Andreev reflection (PCAR) have firmly established that the experimental results of non-magnetic metals ($P=0$, e.g., Cu and Au), and a true half metal ($P=1$ in CrO$_2$) \cite{Ji5} are $quantitatively$ close to those described by the BTK \cite{blonder6} model and the Mazin model \cite{mazin7} respectively. A linear model with a linear combination of the BTK ($P = 0$) and the Mazin($P = 1$) results has been widely used but not rigorously justified. The Dynes model \cite{dynes8} for tunneling has been experimentally established even earlier. The CTC model combines the BTK, the Mazin, and the Dynes models, the three experimentally established models.

The key physics that has eluted previous theories but included in the CTC model is the crucial event at the N/S interface. A single electron cannot go through the N/S interface to have AR because it always requires another electron with the proper spin orientation. On the N side, there is a polarized current with an imbalance of spins, whereas on the S side, the supercurrent carries no spin angular momentum. In ARS, charge is always conserved, as guaranteed by the AR, but spin angular momentum is $not$. The spin angular momentum carried by the redundant majority must become evanescent. The AR wavefunction in the CTC model is $\psi_{AR}=ae^{\alpha q^-z}e^{iq^-x}$ , where $z$ in the first exponential is the coordinate of the spin current, generally unrelated to the charge current coordinate $x$ in the second exponential. This is the same situation encountered in optical total reflection on a 2D surface with the spin current and the charge current as the two effective dimensions. However, in 1D ARS, the evanescent spin current is also along the $x$ direction, thus we have the wave function $ae^{(\alpha+i) q^-x}$ and only the derivative boundary condition needs a factor of 2, which has been attributed to the dimensionless interfacial factor Z. The paper \cite{chen1} clearly emphasizes this essential physics, starting from the abstract.

	Unfortunately, the Comment fails to recognize this crucial novelty and erroneously treats the spin current as charge current and calculates the current of the evanescent wave function, which we have emphasized clearly that it does NOT contribute to charge current. The Comment further incorrectly assumes the evanescent wave is in the superconductor ($x>0$ ) while it is actually on the normal metal side with $x<0$. Since it occurs only near the interface, the calculated $j_{AR}\sim e^{2\alpha q^-x}$ by the Comment is zero at $x<<0$  as expected for the evanescent spin current. In the CTC model, charge conservation is guaranteed by the conservation of probability. The charge current density in the normal metal is always $j_{C}= ej_P$ , where $j_P$  is the probability current. Specifically for the AR term, the charge current is $eA$, with $A$ as the $AR$ probability. The CTC model violates no physical laws. Should the CTC model violate any, so would the three models (BTK, Mazin, and Dynes) it encompasses, not to mention disagreement with experiments.

Without the explicit inclusion of spin into the wavefunction, the theoretical ARS analyses cannot determine $P$, which is defined by the number of conduction electrons in the two spin directions. The AR is not a single electron event since it always needs another electron with proper spin orientation. The availability of the other electron with proper spin should $not$ depend on the dimensions of the interface or the wavevectors of the conduction electrons, as the models \cite {zutic3,grein4} suggested by the Comment, but rather $only$ the spin polarization as represented by $\alpha$ in the CTC model. Therefore, $\alpha$ is $unrelated$ to any characteristic length of charge decay as the Comment speculated. It rather describes the characteristics of the redundant majority spins at the interface. Any model of ARS must be consistent with the two obvious facts: all electrons can go through the interface for $P = 0$  but no electrons go through the interface for $P = 1$, and there should be no variation in spin current on the normal metal side for either case. Indeed, in the CTC model the characteristic length, $\sim 1/\alpha $, is zero for $P = 1$  and infinite with zero magnitude for $P = 0$, so the CTC model $does$ capture these crucial features. Furthermore, for $0<P<1$, the CTC model indicates that the characteristic length is finite but depends on the $P$ value, as expected because the effect of the superconductor should be localized near the interface.  These effects have never been addressed before.

	Because of the constraints from experimental ARS results, we now know any viable theoretical model must provide $quantitative$ results that are close to those of the BTK and the Mazin models at the two limits, or will not be confirmed experimentally. There are indeed many theoretical models but some have serious shortcomings. The Comment goes into great length describing the salient features and merit of some theoretical models \cite{zutic3, grein4}, but neglects to mention that these models do $not$ generate results that are even close to those of experiments (or the BTK and the Mazin models), and some predicted features that have never been observed. For example, the Comment cites the theoretical model of Zutic et al. \cite{zutic3}. While in search of a viable model for analyses during the earlier times of ARS, we have extensively tested that model, which treats the AR events in analogy to the Snell¡¯s law in optics with the suppression of conductance depending on the injection angle. However, after numerous measurements using tips of various materials, shapes, and contact angles, we could not obtain experimental results in any contacts at any temperature that agree with the theory. We now know since that theory does not produce results close to those at the $P = 1$ limit, any experimental attempt for confirmation would be futile. The Comment mentions another more recent model by Grein et al., \cite{grein4}. Among other features, this complex model predicts subgap Andreev bound states for spin-polarized N/S interfaces. Numerous ARS measurements during the last two decades on magnetic materials, including half metals, at various temperatures, some down to less than $1$ K, have never uncovered such subgap features.

	The CTC model provides in concise analytical forms all the probabilities involved in AR with any $P$ value and the results encompass those of the BTK and the Mazin models in appropriate limits as shown in Table 1 of Ref. 1. One immediately knows that the charge current is conserved since for $\alpha = 100 ~(P\cong 1)$, the AR is zero, $A=0$, but the normal reflection is 1 ($B=1$), as shown in Fig. 1 of Ref. 1. Other effects, such as the inelastic scattering in the Dynes model are now included for each probability of any $P$ value. It verifies the validity of the widely used linear polarization model but only under appropriate conditions. These are the key results that no previous theoretical models have provided.

	In summary, there are indeed many theoretical models for ARS, some elaborate, complex, and intricate.  However, only those that can be experimentally confirmed are relevant and useful in extracting physically important quantities in ARS. The CTC model with analytic solutions, unifying the BTK, the Mazin, and the Dynes models, has been experimentally and quantitatively established, and revealed new physics about the evanescent spin current at the N/S interface. The alleged unphysical results are entirely due to the mistreatment of the spin current at the N/S interface and careless calculations in the Comment.

T. Y. Chen$^1$ and C. L. Chien$^2$

$^1$ Department of Physics, Arizona State University, Tempe, AZ  85287, $^2$ Department of Physics and Astronomy, Johns Hopkins University, Baltimore MD 21218

We are confident that Z. Tesanovic, who passed away recently, would concur with the content of the Response.


\begin{thebibliography}{99}

  \bibitem{chen1}
    T. Y. ~Chen, Z. ~Tesanovic, and C. L. ~Chien, Phys. Rev. Lett. {\bf 109}, 146602 (2012).

  \bibitem{schrig2}
    M. ~Schrig, A. A. ~Golubuv, I. I. ~Mazin, B. ~Nadgorny, Y. ~Tanaka, O. T. ~Valls, and I. ~Zutic, arXiv.1301.3511.

  \bibitem{zutic3}
    I. ~Zutic and O. T. ~Valls, Phys. Rev. B {\bf 61}, 1555 (2000).

  \bibitem{grein4}
   R. ~Grein, T. ~Lofwander, G. ~Metalidis, and M. ~Eschrig, Phys. Rev. B {\bf 81}, 094508 (2010).

  \bibitem{Ji5}
    Y. ~Ji, G. J. ~Strijkers, F. Y. ~Yang, C. L. ~Chien, J. M. ~Byers, A. ~Anguelouch, G. ~Xiao, and A. ~Gupta, Phys. Rev. Lett. {\bf 86}, 5585 (2001).

  \bibitem{blonder6}
   G. E. ~Blonder and M. ~Tinkham and T. M. ~Klapwijk, Phys. Rev. B {\bf 25}, 4515 (1982).

  \bibitem{mazin7}
    I. I. ~Mazin, A. A. ~Golubov, and B. ~Nadgorny, J. Appl. Phys. {\bf 89}, 7576 (2001).

  \bibitem{dynes8}
    R. C. ~Dynes, J. P. ~Garno, G. B. ~Hertel, and T. P. ~Orlando, Phys. Rev. Lett. {\bf 53}, 2437 (1984).






\end{thebibliography}
\end{document}